\documentclass[aps,pra,twocolumn,floatfix,footinbib,notitlepage,showpacs]{revtex4-1}
\usepackage{graphicx,graphics,times,bm,bbm,bbold,amssymb,amsmath,amsfonts,dsfont,color}
\usepackage[dvipsnames]{xcolor}
\usepackage[english]{babel}
\usepackage[colorlinks=true, allcolors=blue]{hyperref}
\usepackage{changes}
\usepackage{physics}
\newcommand{\imm}{{\rm i}}
\newcommand\MB[1]{{ #1}}
\begin{document}
\title{Metrology of weak quantum perturbations}
\author{Sidali Mohammdi}
\affiliation{Laboratoire de Physique Th\'eorique, Facult\'e des Sciences Exactes,
	Universit\'e de Bejaia, 06000 Bejaia, Algeria}
	
	\author{Matteo Bina}
\affiliation{MIM - USR Lombardia, LSS {\em Albert Einstein}, I-20137 Milano, Italy}
	
	\author{Abdelhakim Gharbi}
\affiliation{ Laboratoire de Physique Th\'eorique, Facult\'e des Sciences Exactes,
	Universit\'e de Bejaia, 06000 Bejaia, Algeria}
	
	\author{Matteo G. A. Paris}
	\affiliation{Quantum Technology Lab, Dipartimento di Fisica 
	{\em Aldo Pontremoli}, Universit\`a degli Studi di Milano, I-20133 
	Milano, Italy 
	and INFN, Sezione di Milano, I-20133 
	Milano, Italy}
\begin{abstract}
We consider quantum systems with a Hamiltonian containing a weak perturbation 
i.e. \MB{$\boldsymbol{H=H_0} + \boldsymbol{\lambda} \cdot \boldsymbol{\tilde{H}}$}, $\boldsymbol{\lambda}= \{\lambda_1, \lambda_2,...\}$,   \MB{$\boldsymbol{\tilde{H}}$} $= \{H_1, H_2,...\}$, $\left|\boldsymbol{\lambda}\right| \ll 1$, and address situations where \MB{$\boldsymbol{\tilde{H}}$} is known but the values of the couplings $\boldsymbol{\lambda}$ are unknown, and should be determined by performing measurements on the system. We consider two scenarios: in the first one \MB{we assume} that measurements are performed on a given stationary state of the 
system, e.g., the ground state, whereas in the second one an initial state is prepared and then measured after evolution. In both cases, we look for the 
optimal measurements to estimate the couplings and evaluate the ultimate 
limits to precision. In particular, we derive general results for one and two couplings, and analyze in details some specific qubit models. Our results indicates that dynamical estimation schemes may provide enhanced precision 
upon a suitable choice of the initial preparation and the interaction time. 
\end{abstract}
\maketitle
\section{Introduction}
It is often the case that relevant physical phenomena correspond to 
weak perturbations to a stable unperturbed situation. This happens 
in a wide range of disciplines, ranging from applied mathematics 
\cite{barry82}, biology~\cite{Hek2010} to chemistry~\cite{MCWEENY1968567} 
and physics \cite{Them2015}. In these situations, the nature of the 
perturbations is usually known, whereas the strenghts of the perturbations 
are the quantities of interest. The Hamiltonian of those systems 
may be generally written as 
\begin{align}\label{genh}
\boldsymbol{H=H_0} + \boldsymbol{\lambda} \cdot \boldsymbol{\tilde{H}}\,,
\end{align}
where $\boldsymbol{H_0}$ and $\boldsymbol{\tilde{H}}= \{H_1, H_2,...\}$ 
are known Hamiltonian operators and  $\boldsymbol{\lambda}= \{\lambda_1, \lambda_2,...\}$ with $\left|\boldsymbol{\lambda}\right| \ll 1$ is a 
vector of {\em small} unknown coupling parameters, whose values are unknown, 
and should be determined by performing measurements on the system. 
To achieve this goal, there are two paradigmatic approaches, which will be referred to as {\em static} and {\em dynamical} estimation schemes 
throughout the paper. In the first one, the system may be prepared 
in a given stationary state, usually the ground state, which is 
measured to gain information about the value of the parameters. 
In a dynamical scenario, the system is instead prepared in a certain 
state, left to evolve for a given interaction time and finally 
measured. In a dynamical estimation scheme, the initial state, 
as well as interaction time, may be optimized and thus the overall 
precision may be enhanced compared to a static scheme, though the 
practical implementation may be more challenging. Besides the case 
of small perturbations to a given Hamiltonian $\boldsymbol{H_0}$, 
the Hamiltonian in Eq. (\ref{genh}) may also describe systems 
where the couplings have some target values $\overline{\boldsymbol\lambda}_0$ 
and the scope of the measurement is to monitor the system \cite{pintos23,aguilar17,ottoMon21}, i.e. to estimate possible deviations $\boldsymbol\lambda = \boldsymbol\lambda_0 - \overline{\boldsymbol\lambda}_0$ from those values.

A convenient framework to investigate the precision achievable by static 
and dynamical estimation schemes is that of quantum estimation theory
\cite{brody98,akio94,lqe1,lqe2,lqe3,lqe4}, which provides a set of tools to determine the measurement that has to be performed on the system, i.e. to  find the observable that is most sensitive to tiny variations of the parameters \cite{alex07,mariona,giorda08,ade09,ee10,sun11,spara15,torres16,hamza19,PhysRevResearch.3.033250,PhysRevResearch.4.033060,victor22}, and to optimize the initial preparation of the probe \cite{cla14,luis15,analia16,rossi15,sone17,bina18,cosco17,sholeh19,roura22,abbas22}.

In particular, if the value of a single parameter is encoded in the family of quantum states $\{|\psi_\lambda\rangle\}$ (usually referred to as the {\em quantum statistical model}), one may prove that the ultimate precision achievable in estimating $\lambda$ is obtained by measuring the observable $L_\lambda$, known as symmetric logarithmic derivative (SLD), 
which is the self-adjoint operator given by 
\begin{equation}\label{SLD_pure}
L_\lambda=2\big[\ket{\partial_{\lambda}\psi_\lambda}\bra{\psi_\lambda}+\ket{\psi_\lambda}\bra{\partial_{\lambda}\psi_\lambda}\big] \, .
\end{equation}
Upon collecting the result of $M$ repeated measurements on identical preparations 
of the system and suitably processing data (e.g. by maxlik \cite{slocumb90,sam92,zde97,kon99}or Bayesian analysis \cite{berihu09,ste09}) the uncertainty in the determination of $\lambda$, i.e., the precision of the estimation scheme, is given by 
\begin{equation}
\hbox{Var} \lambda \simeq \frac{1}{M\, H(\lambda)}
\end{equation}
where $H(\lambda)$ is the so-called Quantum Fisher information of the quantum statistical model $\{|\psi_\lambda\rangle\}$, i.e., 
\begin{equation}\label{QFI_pure}
Q_\lambda=4\left[ \bra{\partial_\lambda\psi_\lambda}\ket{\partial_\lambda\psi_\lambda} - \left| \bra{\partial_\lambda\psi_\lambda}\ket{\psi_\lambda} \right|^2 \right] \, , 
\end{equation}
(notice that $\bra{\psi_\lambda}\ket{\partial_\lambda\psi_\lambda}$ is a 
purely imaginary c-number, i.e., 
$
\bra{\psi_\lambda}\ket{\partial_\lambda\psi_\lambda}^*=\bra{\partial_\lambda\psi_\lambda}\ket{\psi_\lambda}=-\bra{\psi_\lambda}\ket{\partial_\lambda\psi_\lambda}$).

The generalization to the estimation of more than one parameter 
can be obtained by introducing the so-called quantum Fisher information 
matrix (QFIM) $\boldsymbol{Q}$, which is a real symmetric
matrix with  entries 
\begin{equation}\label{QFImatrix_Pure}
Q_{\mu\nu} = 4\left[ \Re \bra{\partial_\mu\psi}\ket{\partial_\nu\psi}  + \bra{\partial_\mu\psi}\ket{\psi}\bra{\partial_\nu\psi}\ket{\psi} \right] \, 
\end{equation}
The QFIM provides a bound on the covariance matrix (CM) of the estimates 
\begin{equation}
\hbox{Cov}(\boldsymbol\lambda)\geq\frac{1}{M}\,\boldsymbol{Q}^{-1}\,.
\end{equation}
This is a matrix inequality, and it cannot, in general, be saturated. 
Physically, this corresponds to the unavoidable quantum noise that 
originate when the SLDs corresponding to different parameters 
do not commute \footnote{Notice that we assumed the parameters 
to be independent, otherwise the QFIM is singular and the 
maximum number of parameters that can be jointly estimated is 
equal to the rank of $\boldsymbol{Q}$.}.
In those cases, the total variance $\sum_{\mu}V(\lambda_{\mu})$ 
(or a weighted combination of the CM elements)
is a more interesting quantity to study, and since the $\mu$th diagonal 
entry of the covariance matrix is just the variance of the parameter 
$\lambda_{\mu}$, the bound on the total variance is given as
\begin{equation}\label{MultiCR}
\sum_{\mu}V(\lambda_{\mu})\geq\dfrac{B}{M} \qquad B=
\text{Tr}\left[\boldsymbol{Q}^{-1}\right]\, .
\end{equation}
The incompatibility between the parameters can be quantified by 
the so-called {\em asymptotic incompatibility} \cite{carollo19,sholeh20,belliardo21,aleasy21}, 
also referred to as the {\em quantumness} of the quantum statistical model. 
This is defined as 
\begin{equation}
R:=\Vert i\,  \boldsymbol{Q}^{-1}\boldsymbol{D}\Vert_{\infty}
\end{equation}
where $\Vert \boldsymbol{A} \Vert_{\infty}$ is the largest eigenvalue of the matrix $\boldsymbol{A}$ and 
\begin{align}D_{\mu\nu}& =-\frac{i}{2}\,\langle\psi_{\boldsymbol \lambda} | 
[L_{\mu},L_{\nu}]|\psi_{\boldsymbol \lambda}\rangle = 
4\Im \bra{\partial_\mu\psi}\ket{\partial_\nu\psi} 
\end{align}
is the  Uhlmann curvature of the statistical model. The quantity $R$ is a real 
number in the range $0\leq R \leq 1$ with the equality $R=0$ satisfied for compatible parameters, i.e., when $\langle\psi_{\boldsymbol \lambda} | 
[L_{\mu},L_{\nu}]|\psi_{\boldsymbol \lambda}\rangle =0$. 
For just two parameters, one may write 
\begin{equation}\label{R}
    R=\sqrt{\frac{\det\boldsymbol{D}}{\det\boldsymbol{Q}}} \, .
\end{equation}
A tighter scalar bound, known as the Holevo-Cramer-Rao bound 
$\sum_{\mu}V(\lambda_{\mu})\geq C_H/M$, with $C_H \geq B$, may also be derived (see \cite{perspective} for details), and the quantumness $R$ provides a bound to the normalized difference between $C_H$ and $B$ as follows
\begin{equation}\label{rbound}
\frac{C_H-B}{B} \leq R\,.
\end{equation} 

In the following Sections, we aim at finding general formulas for $B$ and 
$R$ for estimation problems involving the parameters of weakly perturbed 
systems in both the static and the dynamical estimation scenarios. We 
also analyze some specific models involving qubits, qutrits and 
harmonic oscillators, and where the Holevo-Cramer-Rao bound is 
known analitically, we check whether the inequality in Eq. (\ref{rbound}) 
is tight. More precisely, Section \ref{Sec_Static} is devoted to static 
estimation schemes, with Section \ref{QET_multiparam} reporting general 
results and Sections \ref{ss:qubit}, \ref{ss:qutrit}, and \ref{ss:anha} 
devoted to specific models involving qubit, qutrit, and oscillatory systems, respectively. Section \ref{dynamic_prob} is devoted to dynamical 
estimation schemes, with Section \ref{ss:gend} reporting general 
results and Sections \ref{ss:qubd}, \ref{ss:qutd}, and \ref{ss:anhad} 
discussing specific results for qubit, qutrit, and oscillatory systems, 
also comparing the performance of dynamical schemes to that of the 
corresponding static ones. Section \ref{outro} closes the paper 
with some concluding remarks.

\section{Static estimation of weak perturbations}\label{Sec_Static}
In this Section, we address estimation of weak perturbations in 
systems descibed by one- and two-parameter (time-independent) 
Hamiltonians of the form $H=H_0+\lambda H_1$ and $H=H_0+
\lambda_1 H_1+\lambda_2 H_2$. 
In particular, we assume that the system may be prepared in 
a given state (e.g, the ground state) and that repeated measurements 
may be performed on the system. We derive general expressions 
for the QFI $Q$ and the quantumness $R$ and discuss specific models 
involving qubit, qutrit and oscillator systems.

\subsection{General results for one and two parameters}\label{QET_multiparam}
Let us consider a system with Hamiltonian $H=H_0+\lambda H_1$ where
$\lambda\ll 1$. \MB{The $n$-th eigenstate $\ket{\psi_n}$ 
of $H$ may be obtained perturbatively to first-order in $\lambda$} as follows
\begin{equation}\label{1st order psi}
\ket{\psi_n}=\ket{\psi_{n}^{0}}+\lambda\ket{\psi_{n}^{1}}+O(\lambda^{2}),
\end{equation}
where $\ket{\psi_{n}^{0}}$ are eigenstates of $H_{0}$ and $$\ket{\psi_{n}^{1}}=\sum_{m\neq n}\dfrac{\bra{\psi_{m}^{0}}H_{1}\ket{\psi_{n}^{0}}}{E_{n}^{0}-E_{m}^{0}}\ket{\psi_{m}^{0}}$$ {is the first-order correction to the 
$n$-th eigenstate}. 
\MB{In general $\braket{\psi^{1}}{\psi^{1}}=N\neq 1$, and it is thus 
convenient to introduce the state $\ket{\psi^{1}}=\sqrt{N}\ket{\phi^{1}}$ 
and write the first-order corrected eigenstate $\ket{\psi}$ as a 
combination of two orthonormal states $\ket{\psi^{0}}$ and $\ket{\phi^{1}}$. The subscript $n$ is omitted to simplify notation. The perturbed state 
and its derivative are thus given by
\begin{subequations}\begin{align}\label{psi_n single}
\ket{\psi}&=\ket{\psi^{0}}+\lambda\sqrt{N} \ket{\phi^{1}} \\
\ket{\partial_\lambda \psi}&=\sqrt{N} \ket{\phi^{1}} \, .
\end{align}\end{subequations}}

According to Eq. (\ref{SLD_pure}), the SLD of this general model 
may be written, up to first order in $\lambda$ as 
\begin{equation}\label{SLD_SingleParam}
L_\lambda=\MB{2\sqrt{N}\big[\ket{\psi^{0}}\!\bra{\phi^{1}}\!+\!\ket{\phi^{1}}\!\bra{\psi^{0}}\!+\!2\lambda\sqrt{N} \ket{\phi^{1}}\!\bra{\phi^{1}}\big]}\, ,
\end{equation}
and the corresponding QFI as 
\begin{equation}\label{QFI_SingleParam}
Q(\lambda)=4N+O(\lambda^{2}) \, .
\end{equation}
\MB{The QFI is independent on the perturbation  (up to second-order) 
and proportional to the norm of the first-order correction 
$\ket{\psi^{1}}$}. This is a remarkably intuitive results, linking 
the estimability of a perturbation to its physical effect on the system.  
The same result may be also obtained expressing the QFI in terms of fidelity \cite{liu14} 
\footnote{One has $Q(\lambda) = 8 \lim_{\epsilon\rightarrow0} [1-|\langle\psi_{\lambda-\epsilon/2}|\psi_{\lambda+\epsilon/2}\rangle|^2]/\epsilon^2$ and $|\langle\psi_{\lambda-\epsilon/2}|\psi_{\lambda+\epsilon/2}\rangle|^2=1-\frac12 N \epsilon^2 + O(\lambda) + O(\epsilon^2)$ }.
Notice also that the $\lambda$-dependent term in the SLD leads to negligible 
(second order) contributions to the QFI and may be  dropped. The optimal 
measurement is thus given by 
\begin{equation}
L=2\sqrt{N}\begin{pmatrix}
0&1\\1&0
\end{pmatrix}\,.
\end{equation}
This expression makes it clear that the optimal measurements set coincides 
with the Pauli matrix $\sigma_x$ over the basis $\{\ket{\psi^0},\ket{\phi^1}\}$, i.e. a detection scheme that senses the coherence of the perturbed state in that  basis.
\par
Let us now address the case of systems with Hamiltonian of the form $H=H_{0}+\lambda_{1}H_{1}+\lambda_{2}H_{2}$ where $H_{1}$ and $H_{2}$ are in general 
non commuting operators, $[H_{1},H_{2}]\neq0$. In this case the pertubations 
depend in a non trivial way on two different parameters $\lambda_{1}$ and $\lambda_{2}$, which should be jointly estimated.
For weak perturbations, the \MB{the $n$-th eigenstate of $H$} $\ket{\psi_n}$ 
may be written, in terms of the eigenbasis of $H_{0}$, as follows
\begin{equation}\label{psi_n double}\begin{split}
\ket{\psi_n}=&\ket{\psi^{0}_{n}}+\lambda_{1}\sum_{m\neq{n}}         \frac{\bra{\psi^{0}_{m}}H_{1}\ket{\psi^{0}_{n}}}{E^{0}_{n}-E^{0}_{m}}\ket{\psi^{0}_{m}}\\+&\lambda_{2}\sum_{l\neq{n}}\frac{\bra{\psi^{0}_{l}}H_{2}\ket{\psi^{0}_{n}}}{E^{0}_{n}-E^{0}_{l}}\ket{\psi^{0}_{l}}\\=& \ket{\psi^{0}_{n}}+\lambda_{1}\sqrt{N_1}\ket{\phi_{n,1}^1}+\lambda_{2}\sqrt{N_2}\ket{\phi_{n,2}^1}
\end{split}\end{equation}
where \MB{$\ket{\phi_{n,\mu}^1}$ are states, i.e. the normalized version of the first-order corrections $\ket{\psi_{n,\mu}^1}=\sqrt{N_\mu}\ket{\phi_{n,\mu}^1}$ having squared norms $N_j$ (with $\mu=1,2$ the index of the parameter $\lambda_\mu$).  As we have  done  before, we drop the index $n$ in order to simplify the notation. These states are not} orthogonal one to each other but both are orthogonal to the unperturbed eigenspace of $H_{0}$, hence, we can express the perturbed state 
$\ket{\psi}$ in an orthonormal basis spanned by the triplet  $\{\ket{\psi^{0}},\ket{j},\ket{k}\}$, with $\braket{j}{k}=\delta_{jk}$. Upon writing the states $\ket{\phi^l}\}$ as
\begin{align}
\ket{\phi^1} &=\sin \frac{\theta_1}{2} |j\rangle + e^{i \gamma} \cos\frac {\theta_1}{2} |k\rangle\notag\\
\ket{\phi^2} &=\sin \frac{\theta_2}{2} |j\rangle + e^{i (\gamma+ \varphi)} 
\cos \frac{\theta_2}{2} |k\rangle\notag\,,
\end{align}
the perturbed state 
and its derivatives $\ket{\partial_{\lambda_\mu}\psi}=\ket{\psi_\mu^1}$ may be written
as  \MB{
\begin{align}
&\ket{\psi}=\ket{\psi^{0}}+ \left(\lambda_{1}\sqrt{N_{1}}\cos\frac{\theta_{1}}{2}+\lambda_{2}\sqrt{N_{2}}{\rm e}^{\imm\gamma}\cos\frac{\theta_{2}}{2}\right)\ket{j}
\notag \\ &+
\left(\lambda_{1}\sqrt{N_{1}}\sin\frac{\theta_{1}}{2}+\lambda_{2}\sqrt{N_{2}}{\rm e}^{\imm(\gamma+\varphi)}\sin\frac{\theta_{2}}{2} \right)\ket{k},
\label{PsiPert}
\end{align} 
\begin{align}
\ket{\partial_{\lambda_1}\psi}&= \sqrt{N_{1}}\left ( \cos\frac{\theta_{1}}{2}\ket{j}+\sin\frac{\theta_{1}}{2}\ket{k} \right )\, , \label{Psi1} \\
\ket{\partial_{\lambda_2}\psi}&= \sqrt{N_{2}}\,{\rm e}^{\imm\gamma}\left (\!\cos\frac{\theta_{2}}{2}\ket{j}+ {\rm e}^{\imm\varphi}\sin\frac{\theta_{2}}{2}\ket{k}\!\right )\,. \label{Psi2}
\end{align} 
In order to quantify the orthogonality between the two 
perturbations, we consider the overlap $\omega = \bra{\phi_{1}^1}\ket{\phi_{2}^1}$ between the two first-order corrections, i.e.,
\begin{equation}\label{overlap}
\omega =\cos\frac{\theta_1}{2} \cos\frac{\theta_2}{2}{\rm e}^{\imm\gamma}+\sin\frac{\theta_1}{2} \sin\frac{\theta_2}{2} {\rm e}^{\imm(\gamma+\varphi)}
\end{equation}
The SLD operators $L_1$ and $L_2$ for the two parameters $\lambda_1$ 
and $\lambda_2$ may be calculated according to Eq.~(\ref{SLD_pure}). The explicit
expressions are reported in Appendix \ref{Appendix_A}. The corresponding
QFIM $\boldsymbol{Q}$ and Uhlmann curvature $\boldsymbol{D}$ are given by
\begin{equation}\label{Q_gen}
\boldsymbol{Q}=
\begin{pmatrix}
4 N_1 & 4\sqrt{N_1 N_2} \,\Re\omega \\
4\sqrt{N_1 N_2} \,\Re\omega & 4 N_2  \\
\end{pmatrix} \, ,
\end{equation}
\begin{equation}\label{D_gen}
\boldsymbol{D}=
\begin{pmatrix}
0 & 4\sqrt{N_1 N_2} \, \Im\omega \\
-4\sqrt{N_1 N_2} \, \Im\omega& 0 \\
\end{pmatrix} \, .
\end{equation}
The ultimate bound $B$ and the quantumness $R$ thus read as follows
\begin{equation}\label{B_gen}
B=\frac{N_1+N_2}{4 N_1 N_2\left[1-\Re^2\omega \right]} \, ,
\end{equation}
\begin{equation}\label{R_gen}
    R=\sqrt{\frac{\Im^2\omega}{1-\Re^2\omega}} \, .
\end{equation}
As expected, the overlap between the perturbations is involved in all the 
quantities of interest. In particular, a real overlap ($\Im \omega=0$) always provides maximum compatibility ($R=0$) between the parameters to estimate. Moreover, if the overlap is zero (both $\Re \omega=0$ and $\Im \omega=0$), i.e. perturbations are orthogonal, the QFI matrix is diagonal, meaning that parameters are 
uncorrelated. On the other hand, if the overlap is a just a phase factor, we have
$\Re^2 \omega+\Im^2 \omega=1$, and thus $R=1$, i.e. maximal incompatibility 
between the parameters. This may happen also when the dimension of the probing 
system is insufficient to estimate a certain number of parameters , as 
it will be illustrated in the next Section by means of a qubit statistical model.}

\subsection{Qubit models}\label{ss:qubit}
Let us consider a qubit system described by the orthonormal basis states 
$\{\ket{0},\ket{1}\}$ of the unperturbed Hamiltonian $H_0=\sigma_z$ with eigenenergies $E_0=1$ and $E_1=-1$. The perturbed Hamiltonian is given by $H=\sigma_z+\lambda \sigma_x$, where $\sigma_z$ and $\sigma_x$ are standard Pauli matrices, and $\lambda$ is the small perturbation parameter that we want 
to estimate. The first-order perturbed ground state is given by 
\begin{equation}
\ket{\psi}=\ket{0}+\frac{\lambda}{2}\ket{1} \, ,
\end{equation}
and the first-order corrected state is $\ket{\psi^1}=\frac12 \ket{1}$ with (squared) norm $N=1/4$. The corresponding SLD is $L_\lambda=\sigma_x$ and the QFI is given by 
\begin{eqnarray}
Q&=&1+O(\lambda^2)
\end{eqnarray}
confirming the general results in Eqs. (\ref{QFI_SingleParam}) and 
(\ref{SLD_pure}).

Let us now consider the more interesting case of a two-parameter perturbation, which 
highlights the issues arising from using an under-dimensioned (compared to the number of parameters) probe system. The perturbed Hamiltonian is $$H=\sigma_z + \lambda_1 \sigma_x + \lambda_2 (\cos\alpha\, \sigma_x + \sin\alpha\, \sigma_y)\,,$$ 
where $\lambda_i$ (with $i=1,2$) are the perturbation parameters and $\alpha$ 
denotes a mixing angle which governs the orthogonality of the two perturbations. 
The first-order perturbed ground state of the system is given by 
\begin{equation}
\ket{\psi}=\ket{0}+\frac12\left(\lambda_1+\lambda_2\,{\rm e}^{\imm\alpha}\right)\ket{1} \, .
\end{equation}
Looking at the above equation, it is clear that the two perturbations cannot, in general, generate two orthogonal states where information about the two parameters 
is encoded \cite{sholeh20}. In fact, the first-order corrected states corresponding to $\lambda_1$ and $\lambda_2$ are the same state except for a phase factor. In other words, 
the two perturbations lead to two degenerate states proportional to $\ket{1}$. Referring to the Bloch sphere representation introduced above, we have $\theta_1=\theta_2=0$ and $\gamma=\alpha$. The overlap in (\ref{overlap}) is given by 
$\omega={\rm e}^{\imm\alpha}$ and the QFIM displays off-diagonal elements. In order to make the two parameters compatible, a probe system with larger dimension should be necessarily employed (see the next Section).

\MB{
\subsection{Qutrit models}\label{ss:qutrit}
Let us consider a three-dimensional spin-1 system with a perturbed Hamiltonian 
given by $H=S_z + \lambda_1 S_x + \lambda_2 (\cos\alpha \,S_x + \sin\alpha \, S_y) $, where $\{S_z,S_x,S_y\}$ denote the irreducible representation of spin-1 
operators in the z-basis:
\begin{equation}\begin{split}
S_z=&
\begin{pmatrix}
1 & 0 & 0 \\
0 & 0 & 0  \\
0 & 0 & -1
\end{pmatrix} \, , \,
S_x=\frac{1}{\sqrt{2}}
\begin{pmatrix}
0 & 1 & 0 \\
1 & 0 & 1  \\
0 & 1 & 0 
\end{pmatrix} \, , \\
S_y=&\frac{1}{\sqrt{2}}
\begin{pmatrix}
0 & -\imm & 0 \\
\imm & 0 & -\imm  \\
0 & \imm & 0 
\end{pmatrix} \, ,
\end{split}\end{equation}
with $\ket{m_s}$, $m_1=\{1,0,-1\}$ being the standard eigenvectors and 
eigenvalues of $S_z$. This Hamiltonian is the direct generalization of 
that considered in the previous Section, and a comparison will reveal 
the role of system dimension. 

For the eigenstate $\ket{\psi^0}=\ket{0}$,  the first-order correstions 
are given by  
\begin{subequations}\begin{align}
\ket{\psi_1^1}=\, &\frac{\ket{1,-1}-\ket{1,1}}{\sqrt{2}}=\ket{\phi_1^1} \\
\ket{\psi_2^1}=\, &\frac{{\rm e}^{\imm\alpha}\ket{1,-1}-{\rm e}^{-\imm\alpha}\ket{1,1}}{\sqrt{2}} = \ket{\phi_2^1} \, ,
\end{align}\end{subequations}
with squared norms given by $N_1=N_2=1$. It is easy to see that these 
perturbation states live in two-level subsystem spanned by $\ket{j}=\ket{1}$ and 
$\ket{k}=\ket{-1}$, and that they may be expressed as in Eq. (\ref{PsiPert}) by
setting  $\theta_1=\theta_2=3\pi/2$, $\gamma=-\alpha$ and $\varphi=2\alpha$. The resulting overlap is real and given by $\omega=\cos\alpha$. In this case, the resulting QFI matrix $\boldsymbol{Q}$ and the ultimate bound $B$ are
\begin{subequations}\begin{align}
\boldsymbol{Q}=\, 4
\begin{pmatrix}
1 & \cos\alpha \\
\cos\alpha& 1  \\
\end{pmatrix} \qquad
B=\,\frac{\csc^2\alpha}{2} \, ,
\end{align}\end{subequations}
whereas the mean Uhlmann curvature is vanishing and the quantumness is zero $R=0$. Moreover, the two perturbed states become orthogonal for $\alpha=\pi/2$, which corresponds to apply non-overlapping perturbations. The QFI matrix becomes diagonal, meaning that the two parameters to be estimated are uncorrelated, and the ultimate bound $B=1/2$ is minimal and coincides with the Holevo bound $C_H$. 
Notice that perturbing a different eigenvector, say $\ket{1,1}$, the situation is dramatically different, since  he two perturbations $S_x$ and $S_y$ generate
the same first-order perturbed state $\ket{1,0}$ and the resulting overlap 
is $\omega={\rm e}^{\imm\alpha}$. 

Summarizing, a two-parameter perturbation cannot be suitably characterized (with maximum precision and compatibility) using a qubit system, whereas the use of a qutrit system allows one to achieve the ultimate limits to precision, via a proper choice of the encoding Hamiltonian terms and of the initial unperturbed state.}

\subsection{A quantum anharmonic oscillator model}\label{ss:anha}
An relevant class of models that may be treated in our formalism is
that of a quantum oscillator weakly perturbed by anharmonic terms. Here
the anharmonic couplings are the quantities to be estimated, e.g., because they 
may represent a resource \cite{alba16}. The Hilbert space of the system is infinite dimensional and may offer an ideal playground to encode as much information 
as needed. 

For the sake of simplicity, we choose natural units ($\hbar=1$) 
and set the frequency and the mass of the oscillator to one $m=\omega=1$. 
The position and momentum operators can be expressed in terms of ladder 
operators as $x=(a+a^{\dagger})/\sqrt{2}$ and $p=i(a^{\dagger}-a)/\sqrt{2}$ 
and hence the unperturbed hamiltonian can be expressed as $H_0 = (p^2 +x^2)/2 
=a^{\dagger}a +1/2$ with corresponding eigenenergies $E_n^0=n+1/2$.
We consider anharmonic perturbations to the harmonic potential such that 
the perturbed Hamiltonians read 
\begin{equation}\label{Anharmonic_Pert}
H= \frac12 \left( p^2 + x^2\right) + \epsilon_1 x^3 + \epsilon_2 x^4,
\end{equation}
where we introduced the two anharmonic couplings $\epsilon_1$ and 
$\epsilon_2$ as the unknown parameters to be estimated. 
The perturbed ground state of the system may be obtained as 
in Eq. (\ref{psi_n double}), where the two first order corrections 
are given by 
\begin{subequations}\label{PertStates_Anharmonic}\begin{align}
\ket{\psi^1_1}&=-\frac12\left( \frac{3}{\sqrt{2}}\ket{1}+ \frac{1}{\sqrt{3}}\ket{3} \right)\\
\ket{\psi^1_2}&=-\frac12\left( \frac{3}{\sqrt{2}}\ket{2}+ \frac{1}{2}\sqrt{\frac{3}{2}}\ket{4} \right) \, ,
\end{align}\end{subequations}
with squared norms $N_1=\frac{29}{24}$ and $N_2=\frac{39}{32}$. The two perturbed states (\ref{PertStates_Anharmonic}) are orthogonal and the same happens 
for any eigenstate of the perturbed Hamiltonian. The QFI (\ref{Q_gen}) 
matrix reads
\begin{equation}
\boldsymbol{Q}=\, 4
\begin{pmatrix}
N_1 & 0 \\
0 & N_2  \\
\end{pmatrix} \, ,
\end{equation}
leading to $B=466/1131\simeq 0.41$. The Uhlmann curvature (\ref{D_gen}) vanishes, corresponding to a zero quantumness (\ref{R_gen}). In conclusion, preparing a quantum oscillator in its vacuum state, is an efficient strategy to precisely sense 
the amplitude of anharmonic perturbations.

\section{Sensing perturbations by dynamical probes}\label{dynamic_prob}
In this Section, we address detection of weak perturbations by performing measurements on an evolved state 
$\ket{\psi(t)} = e^{-\imm H t}\ket{\psi^0}$ where $\ket{\psi^0}$ is a (non stationary) given initial state and $H$ is the
perturbed Hamiltonian under investigation.

\subsection{General results for one and two parameters}\label{ss:gend}
Let us start with a perturbation described by a single parameter Hamiltonian. 
In order to obtain the QFI it is convenient to move into the interaction 
picture (with respect to the unperturbed Hamiltonian $H_0$), where the 
state vector is \MB{given by the unitary transformation} $\ket{\psi_I (t)} 
= U_0^\dag (t) \ket{\psi(t)}$, being $U_0 (t) = e^{-\imm H_0 t}$. The whole time evolution is expressed by 
\begin{subequations}\begin{align}
\ket{\psi_I (t)} & = U_I (t) \ket{\psi^0}\,, \\
U_I (t) & = \mathcal{T} \Big[ \exp{- \imm \lambda K(t)} \Big]\,, \\
K(t) & = \int_0^t\!\! ds\, U_0^\dag(s) H_1 U_0(s)\,, \label{K1_pure_t}
\end{align}\end{subequations}
where $\mathbb{I}$ denotes the identity matrix, $\mathcal{T}[...]$ denotes time-ordering and the operator $K(t)=K^\dag(t)$ is hermitian. Up to first order in $\lambda$ we have
\begin{equation}\label{UI}
   U_I (t) \simeq \mathbb{I} - \imm \lambda K(t)\, ,
\end{equation}
\MB{Going back to the Schröedinger picture the evolved state and its derivative 
with respect to the unknown parameter may be written as}
\begin{subequations}\label{psi_UI}\begin{align}
\ket{\psi_\lambda (t)} & = U_0 (t) \left[\mathbb{I} - \imm \lambda K(t)\right]    \ket{\psi^0} \\
\ket{\partial_\lambda \psi_\lambda (t)} & = -\imm U_0 (t)  K(t) \ket{\psi^0}\,.
\end{align}
\end{subequations}
The leading-order behaviour corresponds to a $\lambda$-independent (zero-th 
order) expression of the QFI 
\begin{equation}\label{QFI_pure_t}
Q(t) = 4 \left[\bra{\psi^0}\! K^2(t)\ket{\psi^0} -
\bra{\psi^0}\! K(t)\ket{\psi^0}^2 \right]    \,.
\end{equation}
Despite it may appear as a rough approximation, this expression of the 
QFI allows us to grab the main features of the dynamical case and to made a 
comparison to the static one.  The QFI in Eq. (\ref{QFI_pure_t}) depends 
on time and is independent on $\lambda$. In other words, for weak perturbations 
the evolution introduces a time dependence, whereas it does not affect 
the covariant nature of the estimation problem. 
\par
Analogously, in the case of a two-parameter Hamiltonian $H=H_{0}+\lambda_{1}H_{1}+\lambda_{2}H_{2}$, the time evolution operator in the interaction picture can be approximated at first-order as $U_{I}(t)\simeq \mathbb{I}-\imm \int_0^t\!ds \,U_{0}^{\dagger}(s)\left(\lambda_{1}H_{1}+\lambda_{2}H_{2} \right)U_{0}$. Upon introducing
the operators 
\begin{align}
K_1(t)&=\int_0^t\! ds \, U_{0}^{\dagger}(s) H_1 U_{0}(s) \, ,\label{K1}\\
K_2(t)&=\int_0^t\! ds\, U_{0}^{\dagger}(s) H_2 U_{0}(s) \, , \label{K2}
\end{align}
the leading order of the elements of the QFI matrix may be evaluated as follows 
\begin{align}\label{QFI_gen_2par_dyn}
Q_{11}&=4\left[\bra{\psi^0}\!K_1^2\!\ket{\psi^0} - \bra{\psi^0}\!K_1\!\ket{\psi^0}^2\right]
\notag \\
Q_{12}&=4\left[\Re\bra{\psi^0}\! K_1 K_2\! \ket{\psi^0} -  \bra{\psi^0}\!K_1\!\ket{\psi^0}\! \bra{\psi ^0}\!K_2\!\ket{\psi^0}\right]\notag \\
Q_{21}&=4\left[\Re \bra{\psi^0} \!K_2 K_1\! \ket{\psi^0} -  \bra{\psi^0}\!K_2\!\ket{\psi^0}\! \bra{\psi ^0}\!K_1\!\ket{\psi^0}\right]
\notag \\
Q_{22}&=4\left[\bra{\psi^0} \!K_2^2\! \ket{\psi^0} - \bra{\psi^0}\!K_2\!\ket{\psi^0}^2 \right]\, ,
\end{align}
where we omitted the time dependence. 
The matrix elements of the Uhlmann curvature are given by 
\begin{equation}
D_{12}=4 \Im \bra{\psi^0} \!K_1 K_2\!\ket{\psi^0} = -D_{21} 
\end{equation}
and the quantumness parameter $R$ reads  as follows
\begin{equation}
R= 4\, \frac{ 
\left|\Im
\bra{\psi^0}K_1 K_2\ket{\psi^0}
\right|}
{
\sqrt{\det \boldsymbol Q}
}\, .
\end{equation}
Now that the general framework has been set, in the following we re-examine some 
of the examples of the previous Sections in order to compare the performance of 
static and dynamical estimation schemes.

\subsection{Qubit models}\label{ss:qubd}
Let us consider a single qubit, initially prepared in generic state $\ket{\psi^0}=\text{cos}(\frac{\theta}{2})\ket{0}+{\rm e}^{\imm\phi}\text{sin}(\frac{\theta}{2})\ket{1}$. Given the results of Section \ref{ss:qubit}, we consider a single-parameter perturbation. The system evolves according to the unitary U = $\exp(-\imm tH)$ where $t$ is the interaction time 
and  $H = \sigma_{3}+\lambda \sigma_{1}$ is the perturbed Hamiltonian with $\lambda$ small.  Using Eqs.~(\ref{K1_pure_t}) and (\ref{QFI_pure_t}) we have
\begin{subequations}\begin{align}
K(t)=&\, {\rm e}^{\imm t}\sin t\ketbra{0}{1}+{\rm e}^{-\imm t}\sin t \ketbra{1}{0}\\
Q(t)=&\,4\sin^2 t \left [1-\cos^2(t+\phi)\sin^2(\theta) \right] \, . \label{QFIqubit_single_dyn}
\end{align}\end{subequations}
In order to compare this result with the QFI obtained in the static case, we set $\ket{0}$ as the initial (unperturbed) state at $t=0$, i.e. $\theta=0$. The dynamical QFI is given by $Q (t)=4\,  \sin^2t $, and achieves a maximum at $t=\pi/2$, where it is four times greater than the corresponding static QFI.



\subsection{Qutrit models}\label{ss:qutd}
We consider the same spin-1 system as in Section \ref{ss:qutrit} and the same Hamiltonian. In order to compare results with the static scenario, we set 
the initial state to  $\ket{\psi^0}=\ket{1,0}$, the QFI and the bound B reads:
\begin{align}
\boldsymbol{Q}&= 16 \sin ^2\frac{t}{2}
\begin{pmatrix}
1 & \cos\alpha  \\
\cos\alpha  & 1  \\
\end{pmatrix} \,,
\qquad \boldsymbol{D}=\, \boldsymbol{0}
\\
\boldsymbol{B}& = \left(8 \sin^2 t/2 \sin^2\alpha\right)^{-1} \qquad
\boldsymbol{R} =  0 \, .
\end{align}
The $D$ matrix and the $R$ parameter vanish, i.e. we have compatibility between the two parameters. The bound $B$ is minimized for orthogonal perturbations $\alpha = \pi/2$, 
and for $t=\pi$. As it happens with qubits, in the dynamical scenario the 
bound is improved by a factor four.

\subsection{Anharmonic oscillator}\label{ss:anhad}
We consider the same system of Section \ref{ss:anha}, prepare the oscillator in the unperturbed ground state and let it evolve according to the perturbed Hamiltonian. 
Using Eqs. (\ref{K1}) and (\ref{K2}) we evaluate the QFIM, which is a diagonal 
matrix with entries (see Appendix \ref{appB} for details)
\begin{align}
Q_{11} & = \frac{29}{3} - 9 \cos t - \frac23 \cos 3 t \\
Q_{22} & = 3  \,(7 + \cos 2 t)\,\sin^2 t \\
Q_{12} & = Q_{21} = 0\,,
\end{align}
whereas the quantumness $R$ vanishes.

\begin{figure}[h]
	\centering
	\includegraphics[width=0.8\columnwidth]{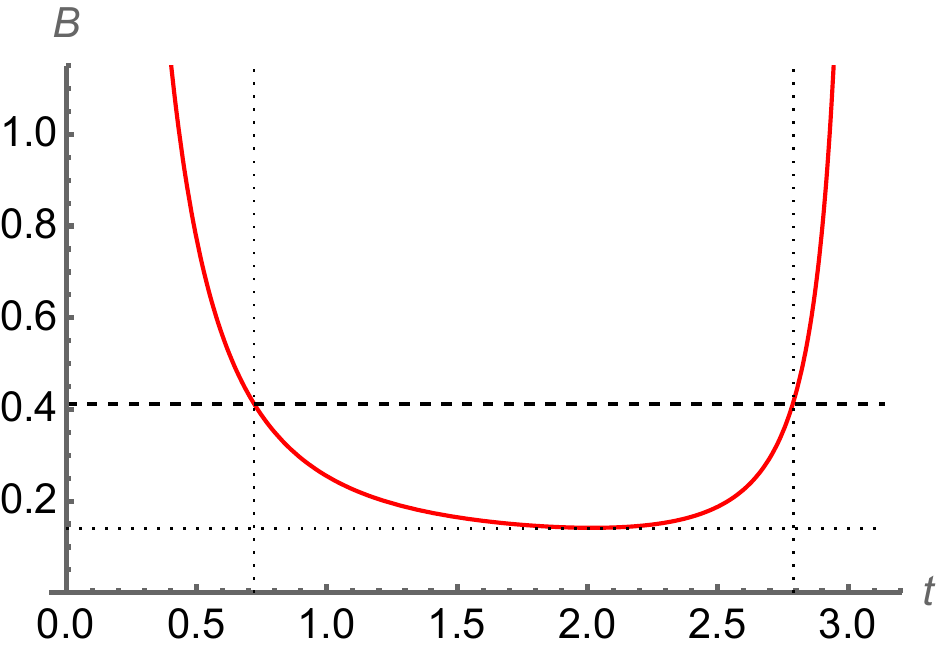}
	\caption{$B$ bound on the total variance for the joint estimation of the anharmonicity parameters as a function of the interaction time. The red solid line is the dynamical bound and the dashed black line denotes the static one. \label{B2}}
\end{figure}

In Fig. \ref{B2} we show the bound $B$ as a function of time ($B$ is a periodic function) compared to the static bound. As it is apparent from the plot, the dynamical scheme beats the static one in the range $t \in (0.721,2.79)$. 
The absolute minimum is obtained for $t \simeq 2.0$, where we have $B\simeq 0.14$, clearly lower than the corresponding static value.

We conclude that preparing the oscillator in the unperturbed ground state and 
performing measurements after a relatively short interaction time  is an effective 
way to reveal the presence of anharmonic perturbations and to estimate their 
amplitudes. 

\section{Conclusions}\label{outro}
In this paper, we have addressed the estimation of weak quantum 
perturbations analyzing two estimation scenarios: a static one, 
where the parameters are inferred by performing measurements 
on a stationary state, and a dynamical one, where the system is 
prepared in a suitably optmized initial state and measurements 
are performed after a given interaction time, which itself may 
be optimized to enhance precision. 

We have found general formulas for the relevant quantities to 
assess precision (i.e. the SLD, the QFIM, the scalar bound $B$ 
on the total variance, and the quantumness $R$) up to the 
leading order in the perturbation parameters, and analyzed in 
some details few quantum statistical models involving 
qubit, qutrit and oscillatory systems.

Our results indicate that dynamical estimation schemes generally 
improve precision, although only for specific preparations of the 
system and values of the interaction time. Ultimately, the choice 
between one scheme and the other does depend on the specific features
of the involved system, and on the experimental difficulties related 
to the preparation of the initial state and the modulation of the 
interaction time.  Our results provide solid tools to compare the 
two approaches in a generic situations.
\par

\section*{Acknowledgement}
We thank Andrea Caprotti for discussions in the early stage of this project. 
This work has been partially supported by MUR through project 
PRIN22-RISQUE-2022T25TR3.

\appendix

\section{Explicit expressions of the SLDs for a two-parameter perturbations}\label{Appendix_A}
Starting from the perturbed state in Eq. (\ref{PsiPert}) and its derivatives 
in Eqs.  (\ref{Psi1}-\ref{Psi2}), the matrix elements $\alpha_{jk} = 
[L_1]_{jk}$ of the SLD operator relative to the parameter $\lambda_1$ reads:
\begin{align}
\alpha_{11}&= 0\\
\alpha_{22}&= 4 \left( \lambda _1 N_1 c_1^2 + \lambda _2 \sqrt{N_1 N_2} c_1 c_2 \cos\gamma\right)\\
\alpha_{33}&= 4 \left(\lambda _1 N_1 s_1^2 + \lambda _2 \sqrt{N_1 N_2} s_1 s_2 
\cos(\gamma+\varphi )\right) \\ 
\alpha_{12}&=\alpha_{21}= 2 \sqrt{N_1}c_1\\
\alpha_{13}&= \alpha_{31} = 2 \sqrt{N_1} s_1\\
\alpha_{23}&= \alpha_{32}^* =  4\lambda _1 N_1 c_1 s_1 \\ 
&\qquad\quad+2 \lambda _2 \sqrt{N_1 N_2}  \left(c_1 s_2{\rm e}^{-\imm (\gamma+ \varphi) }+c_2 s_1 {\rm e}^{\imm \gamma }\right) \,, \notag
\end{align}
whereas $\beta_{jk} = 
[L_2]_{jk}$, i.e. those of the SLD operator relative to the parameter $\lambda_2$ are given by
\begin{align}
\beta_{11}&=0 \\
\beta_{22}&= 4 \left( \lambda _2 N_2 c_2^2 + \lambda _1 \sqrt{N_1 N_2} c_1 c_2 \cos\gamma\right)\\
\beta_{33}&= 4 \left( \lambda _2 N_2 s_2^2 + \lambda _1 \sqrt{N_1 N_2} s_1 s_2 \cos(\gamma+\varphi ) \right)\\
\beta_{12}&= \beta_{21} = 2 \sqrt{N_2} c_2 {\rm e}^{-\imm \gamma }\\
\beta_{13}&= \beta_{31}^* =  2 \sqrt{N_2} s_2 {\rm e}^{-\imm (\gamma+\varphi) } \\
\beta_{23}&= \beta_{32}^*= 4\lambda _2 N_2 c_2 s_2 {\rm e}^{-\imm \varphi} \\
&\qquad\quad
+2 \lambda _1 \sqrt{N_1 N_2} \left(c_1
   s_2{\rm e}^{-\imm (\gamma+\varphi) }+c_2 s_1 {\rm e}^{\imm \gamma }\right)\notag
\end{align}
where $N_j$ is the squared norm of the perturbation vector $\ket{\psi_{n,j}^1}$, $c_{j}=\cos\frac{\theta_{j}}{2}$, and $s_{j}=\sin\frac{\theta_{j}}{2}$, with $j=1,2$.

\section{$K_1$ and $K_2$ for the anharmonic oscillator}\label{appB}
In this Section, we present the explicit expressions of $K_1$ and $K_2$ 
in Eqs. (\ref{K1}) and (\ref{K2}) and their use in evaluating the elements 
of the QFIM. The calculations are tedious but 
straightforward, upon writing the nonlinear Hamiltonians in normal 
order as follows \cite{wilcox67,wilcox70,Candeloro2021}
\begin{align}
x^n & = \frac{1}{2^{n/2}}(a+a^{\dagger})^n \notag \\
& = \frac{n!}{2^{n/2}} \sum_{k=0}^{[n/2]} \sum_{l=0}^{n-2k} 
\frac{a^{\dag l} a^{n-2k-l}}{2^k\, k!\, l!\, (n-2k-l)!}\,,
\end{align}
where $[n]$ denotes the integer part of $n$. We also use the fact
that for a generic function $f(a, a^\dag)$ of the bosonic operators one has
\begin{align}
e^{i y a^\dag a} f(a, a^\dag) e^{-i y a^\dag a} = f(a e^{-i y}, a^\dag e^{i y})
\end{align}
We thus have
\begin{align}
K_1 =& \sum_{k=0}^{1} \sum_{l=0}^{3-2k}\, \int_0^t\! dy \, e^{-i y (3-2k - 2l)} \notag \\
&\times \frac{3!}{2^{3/2}}  
\frac{a^{\dag l} a^{3-2k-l}}{2^k\, k!\, l!\, (3-2k-l)!}\,, \\
= &\,t \,\sum_{k=0}^{1} \sum_{l=0}^{3-2k} 
 e^{-i \frac{t}{2} (2k+2l-3)} \hbox{sinc} \left[\frac{t}{2} (2k+2l-3)\right]\notag\\ & \times
\frac{3!}{2^{3/2}} 
\frac{a^{\dag l} a^{3-2k-l}}{2^k\, k!\, l!\, (3-2k-l)!}\,, 
\end{align}
and
\begin{align}
K_2 =& \sum_{k=0}^{2} \sum_{l=0}^{4-2k}  
\int_0^t\! dy \, e^{-i y (4-2k -2l)} \notag \\
& \times
\frac{4!}{2^2} 
\frac{a^{\dag l} a^{4-2k-l}}{2^k\, k!\, l!\, (4-2k-l)!}\,,\\
= &\, t \,
\sum_{k=0}^{2} \sum_{l=0}^{4-2k} 
e^{-i \frac{t}{2} (2k+2l-4)} \hbox{sinc} \left[\frac{t}{2} 
(2k+2l-4)\right] \notag \\
& \times
\frac{4!}{2^2} 
\frac{a^{\dag l} a^{4-2k-l}}{2^k\, k!\, l!\, (4-2k-l)!}\,.
\end{align}

If we take the unperturbed ground states (the vacuum state of the harmonic oscillator) we have $\langle 0 | K_1 | 0\rangle = 0 $ and $\langle 0 | K_2 | 0\rangle = \frac34 t$. 

In order to calculate the expectation values 
$\langle 0 | K_1^2 | 0\rangle $, $\langle 0 | K_2^2 | 0\rangle$, and 
$\langle 0 | K_1 K_2 | 0\rangle $ and evaluate the QFIM using Eqs. (\ref{QFI_gen_2par_dyn}) we need to calculate expectations values of the form
$\langle 0 | a^{\dag l'} a^{n' -2k'-l'} a^{\dag l} a^{n-2k-l}| 0\rangle$.
In particular, in order to calculate $\langle 0 | K_1 K_2 | 0\rangle $, we need 
\begin{align}
&\langle 0 | a^{\dag l'} a^{n' -2k'-l'} a^{\dag l} a^{n-2k-l}| 0\rangle \notag \\ 
&=  \delta_{l',0}\delta_{l,n-2k} \langle 0 |  a^{n' -2k'} a^{\dag n-2k} | 0\rangle \notag \\
& = \delta_{l',0}\delta_{l,n-2k} \delta_{k',k+\frac{n'-n}{2}} \sqrt{(n' - 2k')!(n-2k)!} \notag \\
& = 0 \quad \hbox{if } n' = n \pm 1\,.
\end{align}
We conclude that $\langle 0 | K_1 K_2 | 0\rangle =0 $ and the same happens for the quantumness $R$.  To calculate the diagonal elements of the QFIM 
we use 
$$
\langle 0 | a^{\dag l'} a^{n -2k'-l'} a^{\dag l} a^{n-2k-l}| 0\rangle = (n-2k)!\,\delta_{l',0}\delta_{l,n-2k} \delta_{k,k'} \,,
$$
such that 
\begin{align}
Q_{11} & =  \, 4\, \langle 0 | K_1^2 | 0\rangle \notag \\
& = 4 \left(\frac{3!}{2^{3/2-1}}\right)^2 \sum_{k=0}^{1} 
\frac{\sin^2 [\frac{t}{2} (3-2k)]}{(3-2k)^2\, 2^{2k}\, (k!)^2\, (3-2k)!}
\notag \\ 
&=  \, \frac{29}{3} - 9 \cos t - \frac23 \cos 3 t 
\end{align}
and
\begin{align}
Q_{22}  = & \, 4\, \left(\langle 0 | K_2^2 | 0\rangle - \langle 0 | K_2 | 0\rangle^2 \right) \notag \\
= & 4 \left\{\left(\frac{4!}{2^{4/2-1}}\right)^2 
\left[\,
\sum_{k=0}^{1} 
\frac{\sin^2 [\frac{t}{2} (4-2k)]}{(4-2k)^2\, 2^{2k}\, (k!)^2\, (4-2k)!}
\right. \right.\notag \\ 
& \left. +\left. \lim_{k\rightarrow 2}
\frac{\sin^2 [\frac{t}{2} (4-2k)]}{(4-2k)^2\, 2^{2k}\, (k!)^2\, (4-2k)!}\right]
- \left(\frac34 t \right)^2
\right\}
\notag \\ 
= & \, 3  \,(7 + \cos 2 t)\,\sin^2 t  \, 
\end{align}

\bibliographystyle{apsrev4-1}
\bibliography{wqp.bib}

\end{document}